# Can it be detected? A computational protocol for evaluating MOF-metal oxide chemiresistive sensors for early disease detection

Maryam Nurhuda[1], Ken-ichi Otake[1], Susumu Kitagawa[1], Daniel M. Packwood[1]*

1. Institute for Integrated Cell-Material Sciences (iCeMS), Kyoto University, 606-8501, Japan

* Corresponding author (dpackwood@icems.kyoto-u.ac.jp)

**Abstract**

Human breath contains over 3000 volatile organic compounds, abnormal concentrations of which can indicate the presence of certain diseases. Recently, metal-organic framework (MOF)-metal oxide composite materials have been explored for chemiresistive sensor applications, however their ability to detect breath compounds associated with specific diseases remains unknown. In this work, we present a new high-throughput computational protocol for evaluating the sensing ability of MOF-metal oxide towards small organic compounds. This protocol uses a cluster-based method for accelerated structure relaxation, and a combination of binding energies and density-of-states analysis to evaluate sensing ability, the latter measured using Wasserstein distances. We apply this protocol to the case of the MOF-metal oxide composite material NM125-$TiO_2$ and show that it is consistent with previously reported experimental results for this system. We examine the sensing ability of NM125-$TiO_2$ for over 100 human-breath compounds spanning 13 different diseases. Statistical inference then allows us to identifies ones which subsequent experimental efforts should focus on. Overall, this work provides new tools for computational sensor research, while also illustrating how computational materials science can be integrated into the field of preventative medicine.

## 1. Introduction

Noncommunicable diseases such as cancers, diabetes, cardiovascular diseases, are the leading causes of death worldwide, resulting in 41 million deaths annually or 74% of all deaths globally.[1] Early diagnosis is important for effective treatment, however diseases in their early stages usually show mild symptoms which are not very noticeable for the patient.[2] As a result, repeated checkups over long periods of time are often necessary for accurate diagnosis, which can be inconvenient for the patient.[3] In order to overcome these problems, novel approaches to early disease detection are required.

Chemiresistive sensors are a promising alternative to the conventional screening methods currently used by health professionals. A chemiresistive sensor can detect health abnormalities based on electrical resistivity changes on contact with blood, urine, skin secretions, or breath samples. Breath sample analysis is particularly appealing for disease detection, as it can be performed non-invasively and by the patient themselves. Human breath contains around 3000 compounds which are the products of endogenous metabolism[4]. The composition and concentration of these compounds in breath provides a fingerprint of the patient's health condition. This point has been vividly illustrated by studies reporting correlations between breath composition and various disease.[5–7] At present, however, there are few materials which can be used as chemiresistive sensors for breath-based disease detection.

Metal oxides are used for a wide range of chemiresistive sensor applications.[8–10] While metal oxides often exhibit strong chemiresistive properties, they tend to exhibit poor selectivity, making them unable to discriminate between different mixtures of compounds. Besides metal oxides, biosensors designed specifically for compounds in exhaled breath have been mainly based on gold nanoparticles[11–13] and carbon nanotubes[14,15]. Recently, a new type of chemiresistive sensor architecture consisting of metal oxide coated with a metal organic framework (MOF) was proposed by Deng *et al*.[15] They reported the material NM125-$TiO_2$, which consists of the metal oxide $TiO_2$ coated with the MOF NM125 ($NH_2$-MIL-125), and showed that it achieves a significantly higher sensitivity towards nitro-explosive compounds compared to ordinary metal oxides, sensing at concentrations as low as 1 part per quadrillion.[16] Such MOF-metal oxide composite sensors are highly attractive for further applications, because their performance can be optimized by changing the organic linker and metal node components of the MOF coating. While potential chemiresistive sensor applications of MOF-metal oxide composite materials have been widely discussed[17–19] the possibility of using them for disease detection in human breath is yet to be explored.

Computational chemistry provides a relatively cheap way to evaluate potential sensing applications of MOF-metal oxide composite materials and other candidate sensors. Previous works have evaluated chemiresistive sensing properties in terms of analyte adsorption energies, density of states (DOS) changes, Fermi level shifts, molecular orbital characteristics, and charge transfer.[20–23] However, computational studies until now have mostly focused on individual analytes. Chemiresistive disease detection requires a different approach, as the sensing characteristics for mixtures of compounds rather than individual analytes must be examined. The analysis of mixtures of compounds requires the use of high-throughput computation to generate results, which in turn require data science for interpretation.

In this paper, we present a new computational protocol for evaluating the sensing ability of MOF-MO composite materials towards small organic compounds found in human breath. Rather than simulating the exact resistivity change of the sensor upon analyte binding, our protocol provides a quick evaluation of sensing ability using a combination of analyte binding energies and DOS analysis. Our protocol is designed to operate in a high-throughput manner, allowing for the generation of computational data for mixtures of compounds found in human breath. We apply our protocol to evaluate the ability of NM125-$TiO_2$, an established nitroexplosive sensor[12], to sense mixtures of compounds associated with various human diseases. Statistical inference is used to suggest that lung cancer might be detectable using NM125-$TiO_2$. While we demonstrate our protocol for the case of human breath compounds, it can be applied to other types of compounds well. This work therefore provides a platform for high-throughput computational evaluation of small organic compounds in a MOF-MO chemiresistive sensor; it also shows how computational chemistry can facilitate research related to early disease detection and preventative medicine.

The structure of the paper is as follows. Section 2 describes our computational protocol. In Section 3, we validate our protocol by applying it to experimentally tested compounds. We then apply the protocol to case of disease-related compound mixtures found in human breath. Discussion and conclusions are presented in Section 4.

## 2. Method

### 2.1. MOF-MO interface model

We consider the interface between the metal-organic framework (MOF) NM125 (NH2-MIL-125, Figure 1a) and the metal oxide rutile $TiO_2$. In this work, we modeled the NH125-$TiO_2$ interface by combining one unit cell of NM125 with the (100) face of a 5.92 Å-thick rutile slab, as described below. The final interface model is shown in Figure 1b.

The interface model was built in four steps. In the first step, a rutile (100) slab was constructed containing 4 x 4 x 2 (100) unit cells on the surface. The dimensions of the rutile slab were 18.38 x 18.38 x 5.92 Å, which closely matched the lattice constant of the cubic NM125 unit cell (18.7 Å). In the second step, the NM125 unit cell and the rutile slab were brought into contact, after expanding the rutile slab. In the third step, a vacuum region of thickness 16.91 Å to one side of the NM125 component to eliminate interactions arising from interfaces in neighboring simulation cells. In the fourth step, the entire system is then relaxed using density functional theory (DFT) as described in Supporting Note 1. The first three steps were performed using the Atomic Simulation Environment Python library.[24] In the fourth step, this structure was relaxed using DFT to obtain the final interface model. DFT was performed as implemented in FHI-aims[25], using a 1 x 1 x 1 *k*-points grid, "light" basis set defaults, and the local density approximation[26] with TS van der Waals corrections [27] . During relaxation, all Ti and O positions were constrained except for the ones in the top 2.96 Å of the rutile slab.

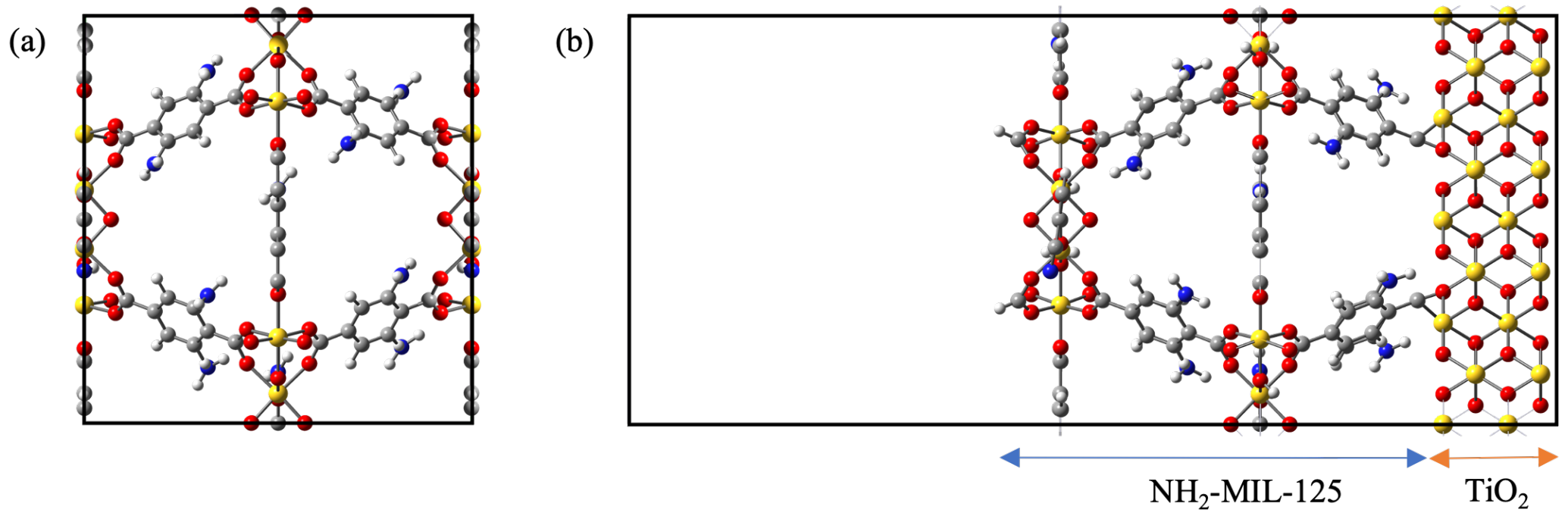

**Figure 1.** (a) A single unit cell of $NH_2$-MIL-125 (NM125). (b) Simulation cell for the NM125-$TiO_2$ interface model. The green boxes represent periodic boundaries. Grey = carbon, white = hydrogen, red = oxygen, blue = nitrogen, gold = titanium.

## 2.2. Cluster model for relaxing analyte compounds

We now consider how analyte molecules can be docked within the interface model described above. Molecule docking on the basis of full structure relaxations are not feasible, especially when applied in a high-throughput manner to large numbers of analyte molecules. The interface model consists of 483 atoms, meaning that standard DFT-based structural relaxation for a single analyte molecule would be in order of days on a typical high-performance server. We therefore employ a cluster-based relaxation method (Figure 2). In this method, a small cluster of atoms is first extracted from the full interface model. The analyte molecule is placed within the cluster, and the analyte and cluster are relaxed together in the gas-phase. The relaxed structure is then mapped back into the full NM125-$TiO_2$ model, from which subsequent property calculations are performed.

To identify a suitable cluster structure, we examined four different candidate clusters and four different analyte molecules (two nitroexplosives RXD and O-DNB, benzene and acetone), as shown in Table 1. Each cluster contains part of the $TiO_2$ slab and different linkers from the original interface model. The four cluster models were evaluated using the following five steps (Figure 2). In step (i), the cluster is extracted from full interface structure by cutting the linkers through the oxygen-metal bonds connecting them to the NM125 metal nodes and passivating any undercoordinated oxygen atoms by adding hydrogen atoms. In step (ii), we placed each molecule in 10 random locations (Supporting Note 2). In step (iii), the molecule and cluster are relaxed for each of the 10 cases using a classical force field. Following relaxation, the binding energy of the analyte molecule, defined as

$$E_b = E_{ca} - \left(E_c + E_a\right), \quad (1)$$

is calculated using the classical force field, where $E_{ca}$ is the energy of the relaxed cluster with analyte included, $E_c$ is the energy of the relaxed cluster without the analyte, and $E_a$ is the energy of the relaxed analyte alone. In step (iv), the case with the lowest energy from the previous step is identified, and DFT is then applied to relax it once more (see Supporting Note 1 for details). In step (v), this relaxed cluster is placed back into the original interface model. This step therefore yields an approximate structure for the relaxed, analyte-docked NH125-$TiO_2$ interface. A final binding energy of the analyte was obtained using a single-point DFT calculation. These binding energies were compared to reference ones computed by directly relaxing the full interface model with analyte molecule included. Step (ii) used the Universal force field[28] from the Forcite module of the Materials Studio software[29]. The Pipeline Pilot software[29] was used to manage the structural relaxations. Steps (iv) and (v) used DFT for relaxation and binding energy calculation, as implemented in the FHI-aims package using the settings described above.

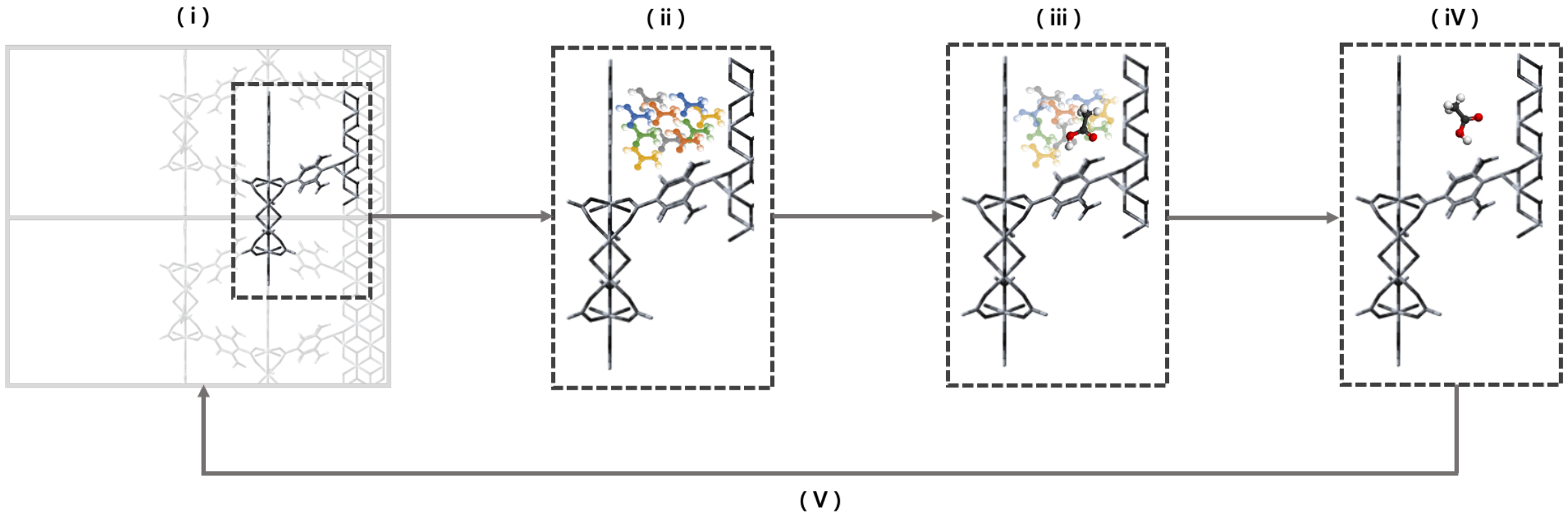

**Figure 2**. Scheme for the cluster approximation approach for geometry optimization. (i) A cluster is cut from the full NM125-$TiO_2$ interface structure (ii) Ten initial positions of each breath compound are generated (indicated by the different colors). Acetone molecules are shown in the figure. (iii) Relaxation with a classical force field is performed for the ten initial positions. (iv) The lowest-energy structure from the previous step is relaxed using density functional theory. (v) The relaxed structure is mapped back into the original NM125-$TiO_2$ interface structure.

The results are summarized in Table 1. The four candidate clusters were built by noting that NM125 contains two types of pores: an octahedral pore and a tetrahedral pore. Cluster model A is a fragment from the octahedral pore, while cluster models B, C, and D are fragments from the tetrahedral pore. The computational times were obtained using servers with 32 Intel

| | Cluster model | | Interaction Energy (eV) | | Computational Time | |
|---|---|---|---|---|---|---|
| | Cluster | Molecule | Full model | Cluster model | Full model | Cluster model |
| A | 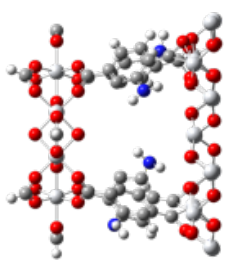 | Acetone | -2.80 | -2.78 | 29:17:51 | 3:41:59 |
| | | Benzene | -2.32 | -2.36 | 28:00:50 | 3:50:31 |
| | | TNP | -9.45 | -9.39 | 35:03:42 | 7:29:09 |
| | | 0-DNB | -5.31 | -5.49 | 19:14:00 | 4:28:13 |
| B | 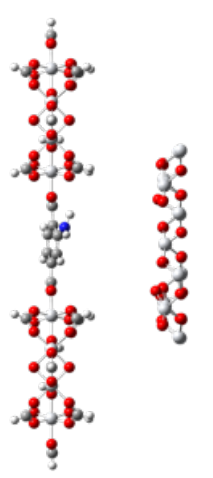 | Acetone | -1.64 | -0.98 | 18:35:48 | 5:02:54 |
| | | Benzene | -1.94 | -0.80 | 14:20:31 | 5:49:23 |
| | | TNP | -6.48 | -2.15 | 30:04:12 | 4:17:33 |
| | | TNT | -6.45 | 93.18* | 35:11:03 | 10:18:26 |
| C | 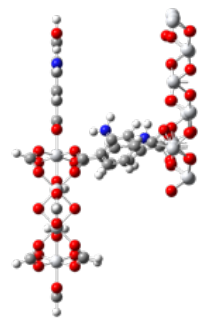 | Acetone | -1.64 | -1.69 | 18:35:48 | 2:49:07 |
| | | Benzene | -1.94 | -1.90 | 14:20:31 | 1:54:27 |
| | | TNP | -6.48 | -5.33 | 30:04:12 | 2:39:16 |
| | | TNT | -6.45 | -5.37 | 35:11:03 | 3:37:31 |
| D | 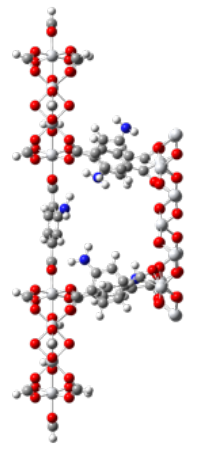 | Acetone | -1.64 | -1.68 | 18:35:48 | 6:34:17 |
| | | Benzene | -1.94 | -1.93 | 14:20:31 | 4:13:04 |
| | | TNP | -6.48 | -6.32 | 30:04:12 | 9:04:38 |
| | | TNT | -6.45 | -6.26 | 35:11:03 | 12:23:15 |

**Table 1.** Comparison of different candidate cluster models. Linkers and nodes in the clusters are highlighted by the pink and yellow boxes, respectively. Computational time is in units of hours:minutes:seconds. See Section 2.2 for details.

Xeon 2.80 GHz cores. All cluster models resulted in a large decrease in computational time compared to the full interface model. Cluster model B resulted quite large (positive) values of the binding energy, and in the case of TNT yielded a very unstable value due to atomic overlaps when mapping the back into the full interface model (indicated by the asterisk). We therefore rejected cluster model B. The remaining cluster models reproduced the binding energies for the weakly adsorbing compounds (acetone and benzene) quite accurately (within 0.05 eV), although less accurately for the strongly binding compounds (TNP, TNT, 0-DNP). Cluster model D was therefore rejected as well due to the large computational times involved, which exceeded 5 hours for two of the molecules involved. The choice of cluster model therefore came down to models A and C. Cluster C achieved much lower computational times than cluster A (less than four hours for each molecule versus), although is less accurate for the cases of strongly interacting compounds. We therefore selected cluster C, restricting our attention to the tetrahedral case and noting that the cluster method is best applied to weakly interacting compounds.

### 2.3. Evaluation of sensor properties

We evaluate the sensor performance of NM125-$TiO_2$ by assessing two properties: binding energy and the extent electronic structure change upon analyte molecule binding. To calculate the binding energy, the compound was first docked into the NM125-$TiO_2$ model using the cluster-based procedure described in the previous section. To measure the extent of electronic structure change, we analyzed the density of states (DOS). The DOS is useful for this purpose, as orbital mixing between the sensor and the analyte will modify the energies and occupancies of the states near the Fermi level and may create new interface states as well. Modifications to the DOS in the vicinity of the Fermi level will affect how easily charges can move through the upper valance bands or lower conduction bands of the sensor, influencing the experimentally observed resistivity change.

Our method for analyzing the DOS involves the following steps (Figure 3). In Step (a), the DOS of the NM125-TiO2 structure was calculated using DFT. This step used the original NM125-$TiO_2$ interface model, without any analyte molecule included. In step (b), we up-weight the DOS in the vicinity of the Fermi level using a Gaussian function. This was performed by computing

$$g(\varepsilon) = f(\varepsilon)\frac{e^{-(\varepsilon-\varepsilon_F)^2/2\sigma^2}}{\sqrt{2\pi\sigma^2}}, \qquad (2)$$

where $\varepsilon$ is the energy, $\varepsilon_F$ is the Fermi energy, $f(\varepsilon)$ is the original DOS as output from the DFT calculation, $\sigma$ is a bandwidth parameter. In this work, we set $\sigma$ = 2 eV for all compounds. Figure 3A shows the original DOS $f(\varepsilon)$ for the NM125-$TiO_2$ structure, the Gaussian function, and the weighted DOS $g(\varepsilon)$, are shown by the black, red, and blue lines, respectively. In step (c), we repeated steps (a) and (b) as above, but this time using the relaxed, analyte-docked NM125-TiO2 model obtained using the cluster method described previously. We denote the resulting DOS for this case was $g_a(\varepsilon)$. Figure 3B shows the $g_a(\varepsilon)$ obtained for the case of acetic acid analyte. Finally, in step (iv), we compare $g$ and $g_a$ using the so-called Wasserstein distance[30]. The Wasserstein distance is a metric used in optimal transport theory, an area probability theory. Intuitively, the Wasserstein distance can be understood as the minimum 'effort' required to transform one probability distribution into another. In terms of Figure 3, the Wasserstein distance would measure the minimum effort required to reshape the blue line in Figure 3A into the one in Figure 3B. The formal definition of the Wasserstein distance is

$$W(g, g_a) = \min_{\pi\in\Pi(g,g_a)} \int |\varepsilon_i - \varepsilon_j| \pi(\varepsilon_i, \varepsilon_j) d\varepsilon_i d\varepsilon_j, \qquad (3)$$

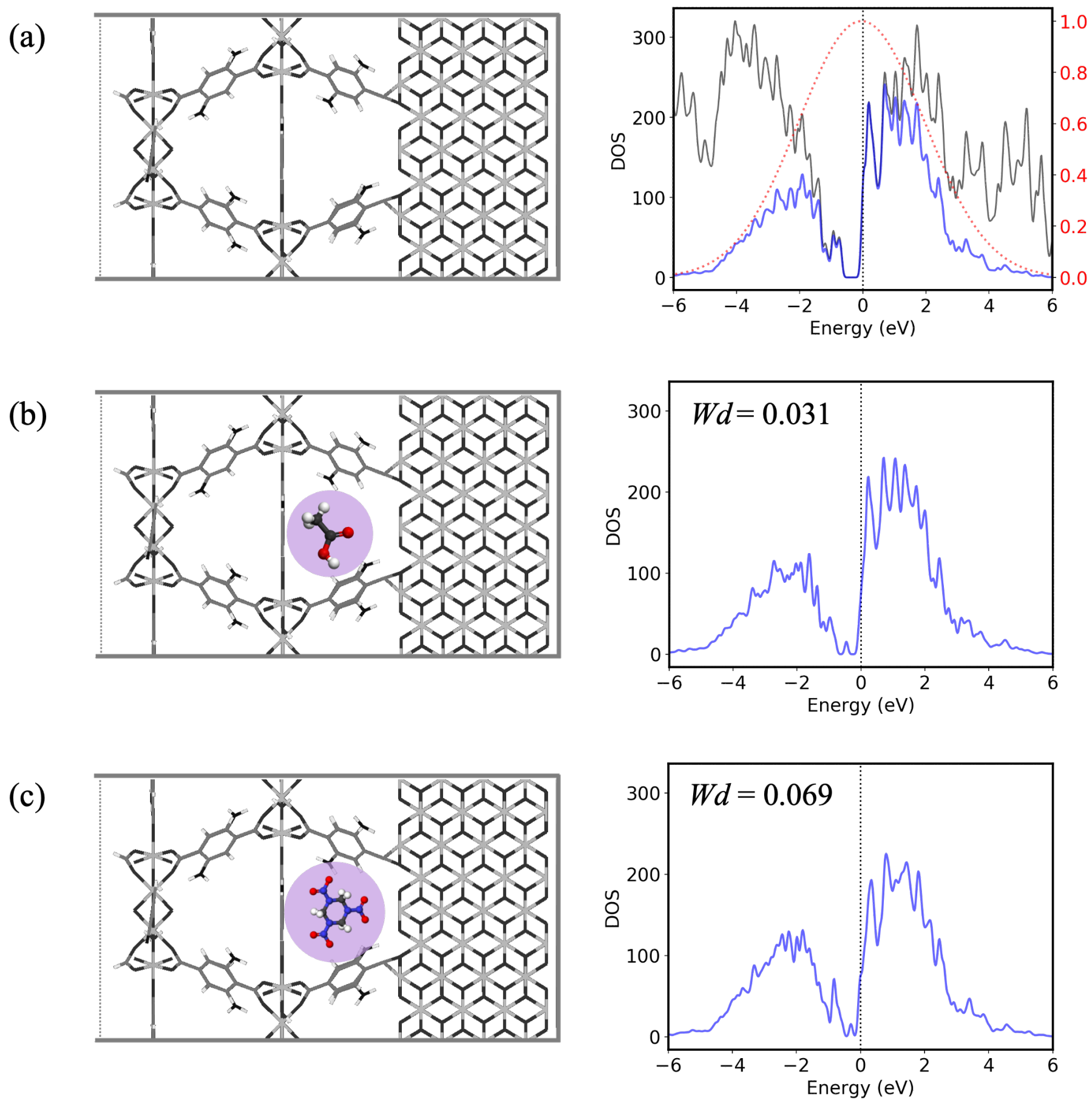

**Figure 3.** Illustration of the Wasserstein distance calculation. (a) Empty NM125-$TiO_2$ interface (left) and density of states (DOS) (right). The black line is the DOS, the red line is the Gaussian function, and the blue line is the Gaussian-weighted DOS. (b) NM125-$TiO_2$ interface model containing the relaxed acetone molecule, as relaxed using the cluster method (left) and the Gaussian-weighted DOS (right). (C) NM125-TiO2 interface model containing the relaxed TNP molecule (left) and the Gaussian-weighted DOS (right).

where $\Pi(g, g_a)$ is the set of all 'couplings' between the $g$ an $g_a$ (see references 31 and 32 for details). We calculated (3) using the Python package SciPy.[33] All DFT calculations were performed as described above.

Figure 3B and Figure 3C show the Gaussian-weighted DOS for the cases of acetic acid- and TNP-docked NH125-$TiO_2$. We obtain $W(g, g_a) = 0.03$ for the acetic acid case, and 0.07 for the TNP case. This indicates that the DOS in the region of the Fermi level undergoes greater changes upon TNP docking compared to acetic acid docking. In both cases, we observe a new peak in the DOS appearing just below the Fermi level at -0.3 eV. This corresponds to a new interface state formed as a result of orbital hybridization between the analyte molecules and the sensor. In the DOS for the bare NH125-$TiO_2$ interface in Figure 1A, we can see a clear two-peak feature at about -1.1 eV. Upon docking of the analyte molecules, this feature undergoes large changes. For both cases, this feature becomes single-peaked. However, for the case of TNP, this peak becomes very sharp and singular in the region. As a result, the Wasserstein distance for TNP case is large compared to the acetic acid case.

### 2.4. Human breath compounds

We considered 124 human breath compounds as listed in Supporting Table 1. These compounds were selected from the Human Breathomics Database[7] and other sources (see Supporting Table 1). These compounds cover a diverse range of chemical types, including acids, alcohols, aldehydes, aromatics, hydrocarbons, and ketones. They are also associated with a broad spectrum of health conditions[7,34,35], including respiratory diseases such as chronic obstructive pulmonary disease (COPD), metabolic conditions such as diabetes, and various cancers such as liver cancer, lung cancer, and gastrointestinal cancer (see Supporting Table 2). The molecular structures for these compounds were retrieved from the PubChem database.[36]

Each breath compound was docked into the NH125-TiO2 model using the cluster method, as described in section 2.2. For each case, we calculated the binding energy and Wasserstein distance using the methods described in section 2.3 above. To validate our method, we also considered the nitro-explosive compounds RDX, 0-DBD, TNP, and TNT, as well as 10 compounds which are known to be undetectable[16]. (acetone, ammonia, benzene, carbon dioxide, carbon monoxide, ethanol, hydrogen sulfide, methane, sulfur dioxide, and toluene; see Supporting Table 3). All calculations were performed in a high-throughput manner using the Pipeline Pilot software[29].

## 3. Results

### 3.1. Validation

We first validate our computational protocol by considering the four nitro-explosive compounds and 10 interfering compounds described above. As shown in Figure 4, the nitro-explosives have strong binding energies in range of around -4 to -6 eV and high Wasserstein distances in range of $3 \times 10^{-2}$ to $7 \times 10^{-2}$, while the interfering compounds have weaker binding energies and show lower Wasserstein distances (full results shown in Supporting Table 3). The calculated binding energy values are overestimated compared to the experimental values[37], however this is an expected due to the use of the local density approximation in our DFT calculations and does not affect binding energy trends. Figure 4 shows that these two groups of compounds can be clearly separated on the basis of binding energy and Wasserstein distance, which confirms that our computational protocol can distinguish between compounds which are strongly and poorly sensed experimentally. We can therefore use these results as a kind of 'calibration curve' and judge whether a compound is likely to be detectable or not based on its proximity to the above points. We will return to this point later.

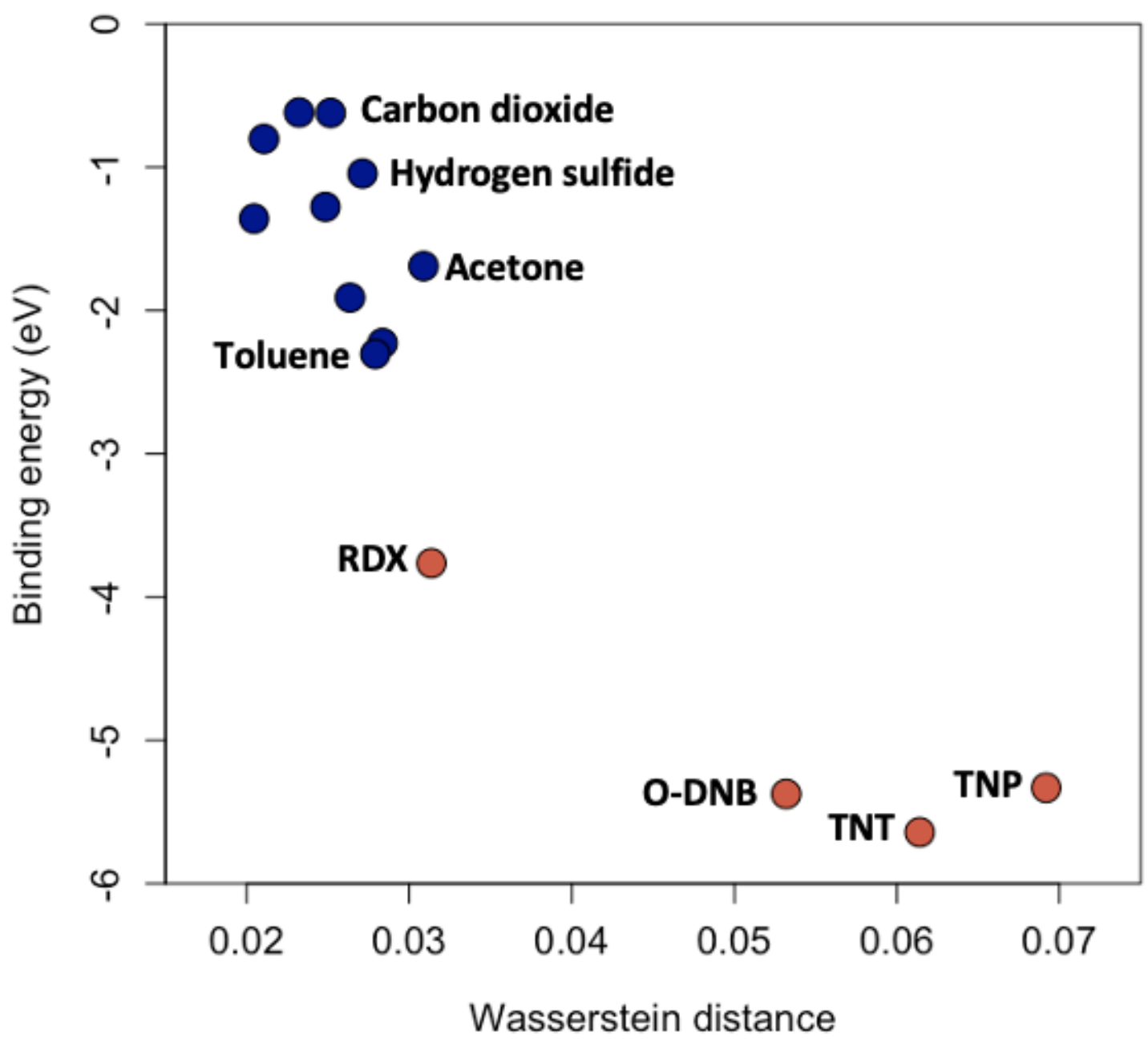

**Figure 4.** Binding energy and Wasserstein distances for nitro-explosives (red points) and interfering compounds (blue points). Only some interfering compounds are labeled for clarity.

### 3.2. Data visualization

In order to visualize the variation in analyte structures in data, we apply principal component analysis (PCA). To perform PCA, we first compute the Coulomb matrix for the 124 breath compounds listed in Table 2, as well as the interfering compounds and nitroexplosives discussed above. The Coulomb matrix for compound $k$ was computed as $\mathbf{C}_k = [C^k_{ij}]_{n\times n}$, where $n$ is the number of atoms and

$$C_{ij}^{k} = \begin{cases} 0.5 Z_i^{2.4} & i = j \\ \dfrac{Z_i Z_j}{R_{ij}} & i \neq j \end{cases}, \tag{4}$$

where $R_{ij}$ is the distance between atoms $i$ and $j$, and $Z_i$ and $Z_j$ are the atomic numbers of atom $i$ and atom $j$, respectively.[38] Coulomb matrices were padded with zeros so that for all cases the dimensions were 58 x 58 (the size of the largest molecule considered). PCA was then performed using the eigenvalue spectra of the Coulomb matrices. Because Coulomb matrix spectra are generic descriptors for molecular geometry, the principal components can be understood as measuring analyte geometry.

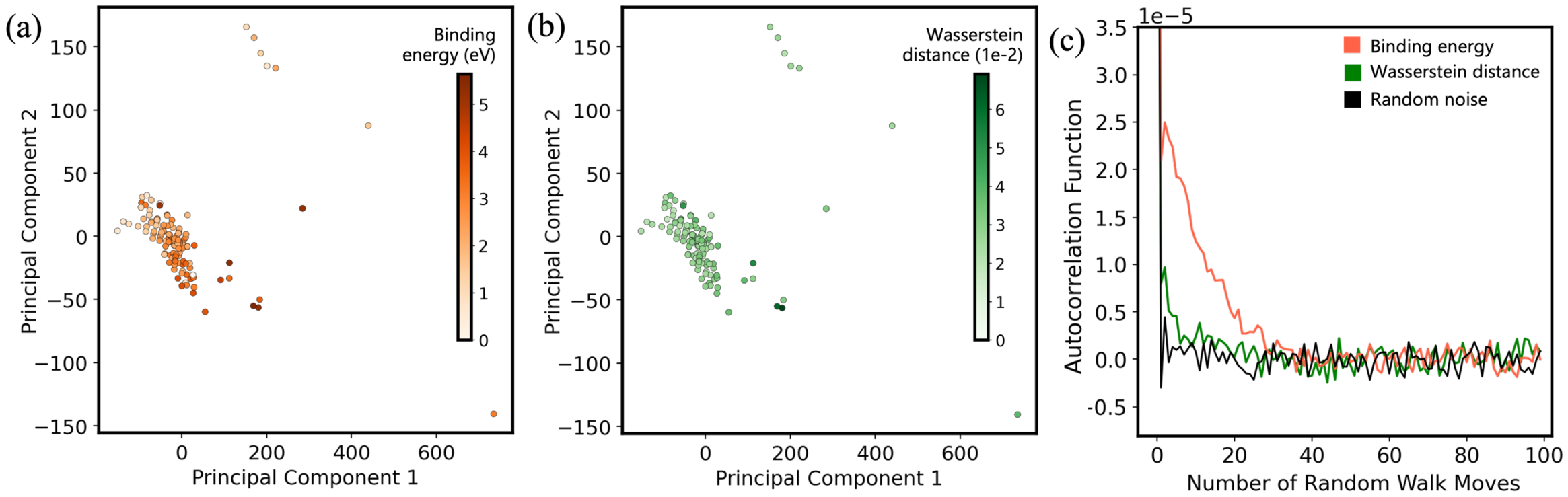

**Figure 5.** Visualisation of calculation results for 124 breath compounds using principal compounds built from Coulomb matrix descriptors.

Figure 5 plots the breath compound data using the first two principal components, and each point corresponds to one compound. Each compound is colored according to its binding energy (Figure 5A) and its Wasserstein distance (Figure 5B). Most compounds are located in a single cluster located on the left-hand side of the plots, with a few others (mainly sulfur- or chlorine-containing compounds, and the strongly binding nitroexplosive compounds) being dispersed away from this cluster. The presence of one large cluster implies that there is little variation in the geometries of the analytes in our database. This is sensible, as the molecules considered mostly quite small and can only adopt a limited number of conformations at equilibrium. While the variation in analyte geometries is small, the spread of this cluster shows that this variation is non-negligible.

We can use Figure 5 to confirm that that binding energies and Wasserstein distances describe different aspects of the analyte-sensor interaction. From a visual inspection it appears that binding energy changes more gradually across this cluster compared to the Wasserstein distance (Figures 5A, B). This suggests that binding energy is correlated more strongly with analyte geometry than the Wasserstein distance is. To rigorously confirm this statement, we simulated a random walk on the points in these plots and computed their autocorrelation functions (ACF)[39]. Details on this calculation are given in Supporting Note 3. The ACFs are shown in Figure 5C. The black curve corresponds to the case where the points are colored according to white noise. This represents a baseline case where a variable is not correlated with the principal components, and hence with analyte geometry. As required, the black curve immediately plummets to zero, indicating negligible correlation. The orange curve represents the ACF obtained when the points are colored according to the binding energy. In clear contrast to the black curve, this curve decays quite slowly, indicating that nearby points in the plot are highly likely to have similar values of the binding energy. These results confirm that binding energies are Wasserstein distances are complementary: binding energies are related by analyte geometry, whereas Wasserstein distances are affected by other factors. The relationship between binding energy and analyte geometry demonstrates that these interactions are a type of host-guest interaction, where the analyte molecule (guest) needs to adopt a particular shape in order to fit within the pore (host). For the compounds studied here, the binding energy therefore appears to be a direct measure of the host-guest interaction strength, in contrast to the Wasserstein distance.

To gain further insight into what the Wasserstein distance is measuring, we performed additional calculations using synthetic DOS curves (Supporting Figure 1). As a baseline case, we considered a DOS consisting of two Gaussians located at equal distances on either side of the Fermi level. These Gaussians represent idealized valance and conduction bands. The DOS for this baseline case was then compared to three cases: (i) where the two Gaussians are shifted slightly towards and away from the Fermi level, (ii) where the Gaussians are widened, and (iii) where third narrow Gaussian is inserted in the middle of them. Thus, case (i) crudely models how the DOS might change as a result of electrostatic factors, whereas case (ii) and (iii) model how it might change as a result of quantum mechanical factors. We find that the Wasserstein distance is quite large when comparing the baseline case to case (iii), providing that the new Gaussian peak is not too narrow. The Wasserstein distance is moderate when comparing to case (i), which suggests some sensitivity to changes in electrostatic environment. On the other hand, the Wasserstein distance is notably smaller when comparing to case (ii), suggesting a reduced sensitivity to orbital mixing. Thus, the Wasserstein distance is mainly affected by the emergence of interface states, and somewhat affected by electrostatic aspects of the analyte-sensor interaction.

### 3.3. Sensitivity to breath compounds and potential disease detection

Figure 6a plots the data computed from the 124 breath compounds. The compounds are colored according to their major chemical classification. Overall, most of the breath compounds overlap with the undetectable compounds: 80 % of them have binding energy weaker (more positive) than -3 eV (Figure 6b), and 90 % of them have Wasserstein distances less than $3.5 \times 10^{-2}$ (Figure 6c). Moreover, we can see that the compounds which reside far from the null region tend to be acidic. These include the acetate ion (binding energy = -5.21 eV, Wasserstein distance = $4.66 \times 10^{-2}$) and the formate ion (-5.1 eV, $4.92 \times 10^{-2}$), both of which can appear in human breath as a result of acetone metabolism. Dodecanoic acid (-4.0 eV, $3.64 \times 10^{-2}$), 1-methylbutylacetate (-3.83 eV, $3.72 \times 10^{-2}$), and ethyl-4-ethoxybenzoate (-4.52 eV, $3.77 \times 10^{-2}$) also lie somewhat away from the undetectable compounds. These compounds contain either free electrons or electrons bound in weak chemical bonds which can easily interact with NM125-$TiO_2$. Interestingly, none of the other chemical classifications dominate the region away from the undetectable compounds. This observation suggests that the NM125-$TiO_2$ sensor will be most effective at identifying diseases based upon acidic or easily ionizable breath compounds.

In Figure 7, we plot the data in Figure 6 again, this time only showing the compound mixtures associated with specific diseases. For reference, we also plot the data for the undetectable compounds. For most diseases considered most of these compounds lie close to the undetectable ones, which strongly suggests that they would be undetectable by NM125-$TiO_2$. However, some diseases are associated with compounds which lie far away from these compounds, such as COPD and lung cancer, suggesting that they might be detectable with this material.

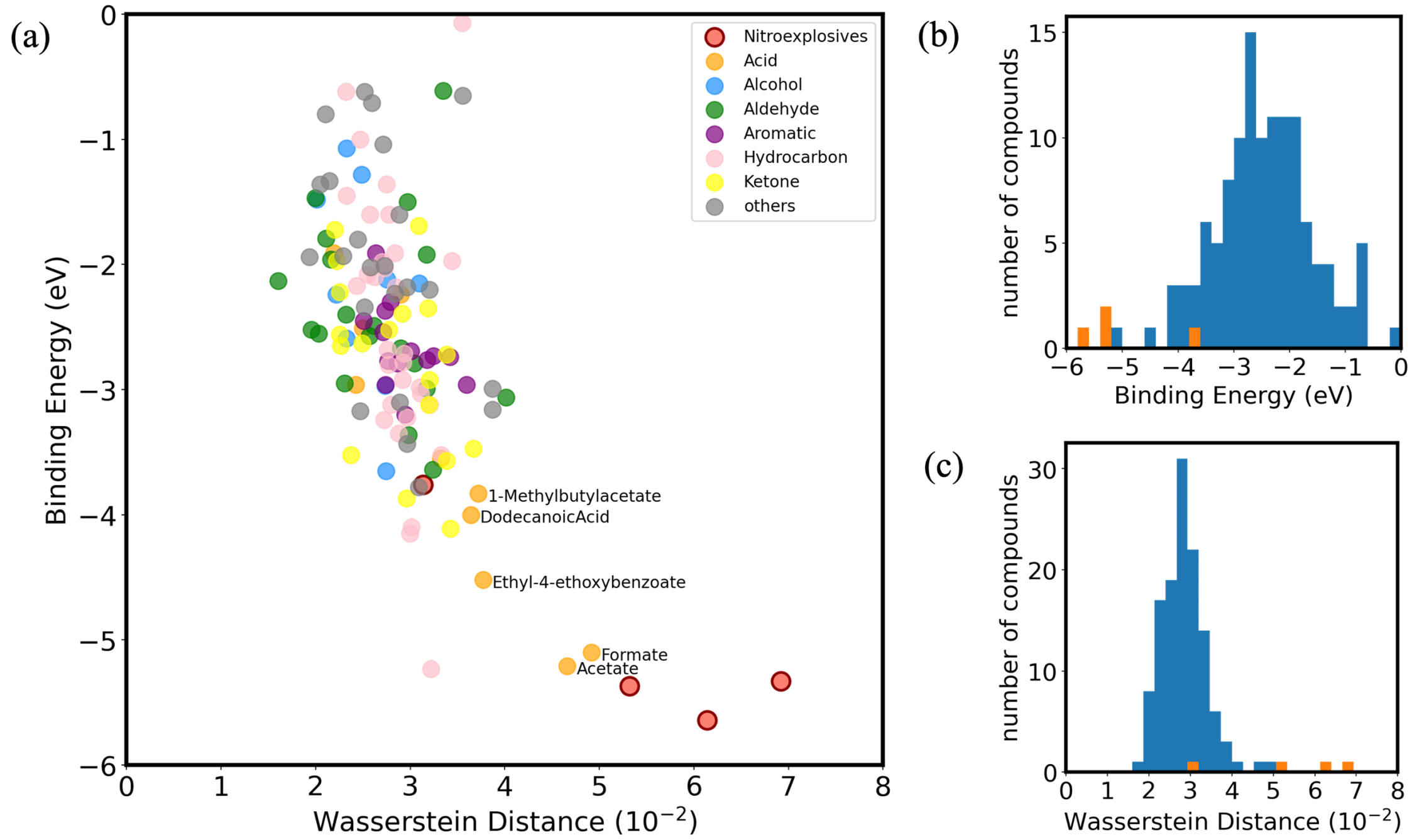

**Figure 6.** (a) Binding energy and Wasserstein distance calculation on 124 breath compounds. For reference, we include and the null region (black part), representing the location of the experimentally undetectable compounds. Compounds are colored by their chemical family, as indicated by the legend. (b) Histrogram of binding energies for the breath compounds (blue) and nitro-explosives (orange). (c) Histogram of Wasserstein distances for the breath compounds (blue) and nitro-explosives (orange).

We can make these statements more rigorous. It is not possible for us to claim definitively that these diseases would be detectable by NM125-$TiO_2$, because we have not showed how binding energies and Wasserstein distances relate to sensor resistance changes. Establishing this connection would require device-level modeling, which goes beyond the scope of this work. However, we can show statistically whether the sensor response to a breath sample from a patient with COPD or lung cancer will differ from that of a mixture of undetectable compounds.

To this end, we compare the distributions of compounds associated with COPD or lung cancer with the distribution of the interfering compounds. Concretely, for each disease, we treat our data as a sample taken from a population representing every compound associated with the disease. Similarly, the data for the undetectable compounds is treated as a sample taken from a population representing every compound that cannot be detected with NH125-

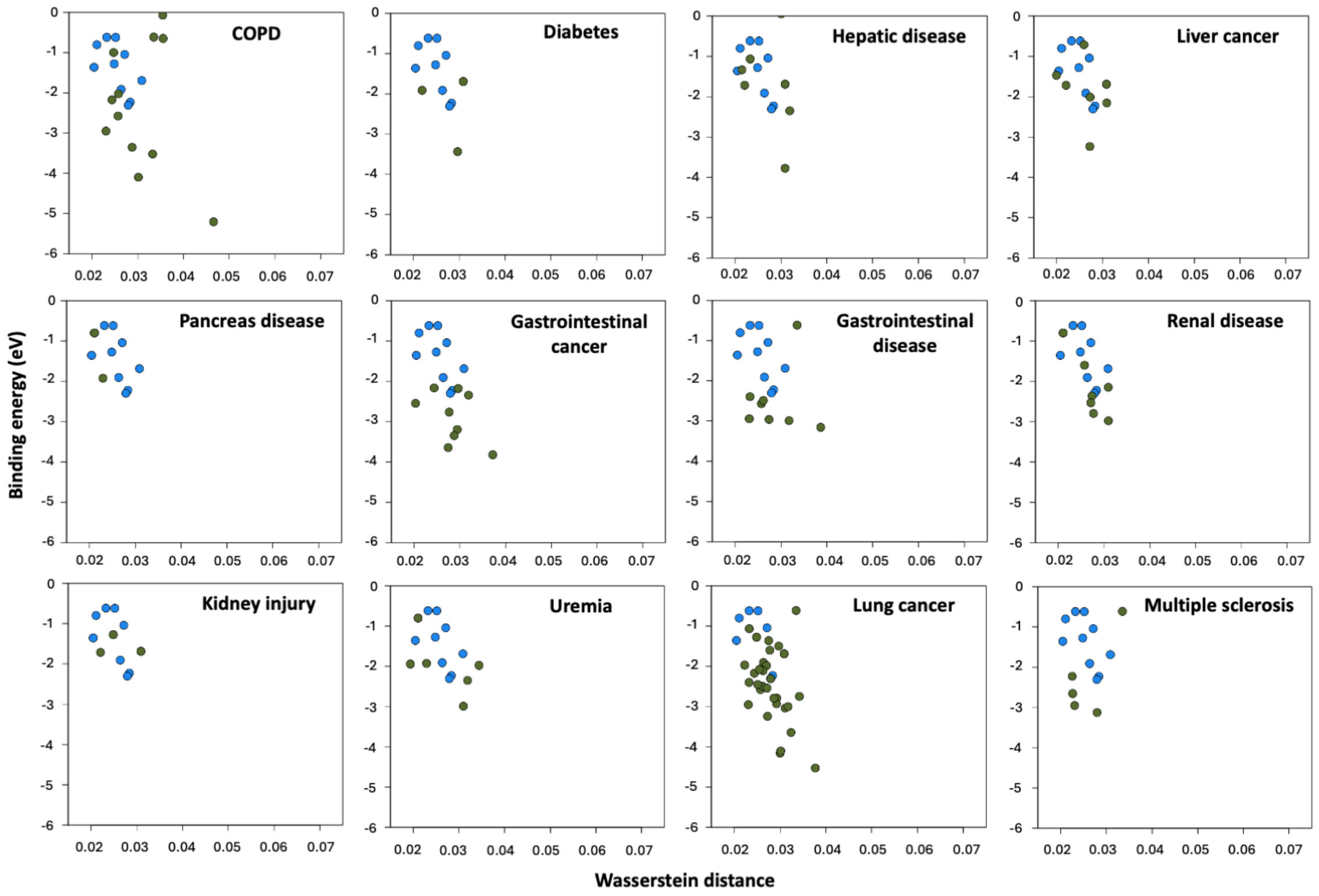

**Figure 7.** As for Figure 6a, but only showing the breath compounds associated with particular diseases. Blue points indicate the undetectable interfering compounds, and green points indicate compounds associated with the disease.

$TiO_2$. For COPD and lung cancer, we then perform a Hotelling test, a method from the field of statistical inference. The Hotelling test considers the null hypothesis that the population average of the vector ($E_b$, $W$) will differ between the disease under consideration and the undetectable compounds. The Hotelling tests were performed using the R software package[40]. Our evaluations are based on the $p$-value, which is the probability of acquiring the sample data if the null hypothesis was true. Following convention, we reject the null hypothesis for $p$-values less than 0.05. We only perform Hotelling tests for two diseases (COPD and lung cancer) in order to reduce the probability of incurring type-1 errors (false rejection of the null hypothesis).

The Hotelling tests yield $p$-values of 0.07 for COPD and 0.006 for lung cancer. Thus, there is insufficient evidence to reject the null hypothesis for COPD, but strong evidence for rejection for the case of lung cancer. This does not confirm that lung cancer would be detectable using NM125-$TiO_2$, but does suggest that a breath sample from a patient with lung cancer would illicit a stronger sensor response than a sample of undetectable compounds.

## 4. Discussion

There are several aspects of our approach which could be targeted for further research. For high-throughput screening applications involving MOF-related materials, structure relaxations will become a serious computational bottleneck. Some kind of approximate procedure is unavoidable for making progress. The cluster method presented in this work accelerates these calculations by isolating a small part of the whole interface structure and performing a series of classical relaxations. We validated the method by comparing to DFT-calculated binding energies obtained from a relaxation of the full structure. However, the method should be validated further by considering how well it can reproduce other properties, particularly ones related to electron exchange and hybridization with the sensor.

In this study, we evaluated sensing ability using in terms of binding energies and Wasserstein distances. These quantities showed clear consistencies with experimental results when computed for experimentally tested compounds, providing a *post-hoc* justification for their use. However, the theoretical justification for using these two quantities, particularly Wasserstein distances, remains somewhat unclear. The sensing ability of a material is determined by a mixture of electrostatic and quantum mechanical factors arising from the molecule-sensor interaction. For the host-guest interaction represented by an analyte within a MOF pore, binding energies should be determined broadly by the geometry of the molecule. Figure 5 confirmed this point, and moreover showed that the Wasserstein distance is independent of molecular geometry. Calculations involving an idealized DOS then showed that the Wasserstein distance was mainly responsive to the appearance of new peaks near the Fermi level, which is an entirely quantum mechanical aspect of the molecule-sensor interaction. Binding energies and Wasserstein distances therefore appear to be complementary, responding to different aspects of the analyte-sensor interaction. However, this discussion remains somewhat speculative. If the Wasserstein distance is to be used in further research, then its relationship with the electrostatic and quantum mechanical aspects of the molecule-sensor interaction should be clarified rigorously from first-principles theory.

In evaluating the sensing ability of NM125-$TiO_2$ for breath-based disease detection we ignored the presence of water molecules. Real breath samples will mainly consist of water vapor, with other compounds dispersed randomly and at much lower concentrations. While it is beyond the scope of this study, future work might consider improving our approach by incorporating the effect of water molecules somehow, perhaps by incorporating them into other pores of the NM125-$TiO_2$ interface model.

This work has only considered NM125-$TiO_2$ as a potential sensor. In doing so, we were able to chart a course showing how computational chemistry can be applied in the problem of early disease detection. However, we did not explore a particularly powerful aspect of modern computational chemistry – the ability of design new materials. Indeed, there range of potential MOF-MO sensors is enormous. Not only are many choices for the metal oxide component possible, but there are a number range of possible linkers and metal ions for the MOF to choose from as well. While NM125-$TiO_2$ appears to be useful for only a couple of human diseases, optimization of the components may result in a far more effective material. An appealing next step is therefore to design an optimal MOF-MO system for early disease detection, perhaps using machine learning methods.

## 5. Conclusions

We have presented a computational protocol for evaluating whether a metal organic framework-metal oxide (MOF-MO) composite material can act as a sensor for small organic molecules present in human breath. This protocol utilizes a new cluster-based method for docking and relaxing molecules at the MOF-MO interface and evaluates sensing properties on the basis of binding energies and Wasserstein distances. The latter quantifies changes near the Fermi level in the density-of-states (DOS) upon molecule binding, and was found to be particularly sensitive to the appearance of interface states. We illustrated our approach by considering the case of the MOF-MO sensor NM125-$TiO_2$, for which experimental results are available to help validate our protocol. We applied our protocol in a high-throughout manner to a sample of over 100 human-breath compounds which are known biomarkers for several diseases, evaluating binding energies and Wasserstein distances for each one. Statistical inference then suggested that a sample of compounds from lung cancer would illicit a stronger sensor response than a sample of undetectable compounds. Subsequent experimental efforts in MOF-MO sensor development should therefore focus on lung cancer as a potential target for disease detection. Overall, our work provides a viable blueprint for integrating computational materials chemistry into the important field of preventative medicine.

## Acknowledgements

This work was supported by JSPS Kiban S (22H05005), the JSPS Bilateral Program (Collaborative Projects), and the Institute for Integrated Cell-Material Sciences (iCeMS).

**Supporting Information for**

# Can it be detected? A computational protocol for evaluating MOF-metal oxide chemiresistive sensors for early disease detection

Maryam Nurhuda[1], Ken-ichi Otake[1], Susumu Kitagawa[1], Daniel M. Packwood[1]*

1. Institute for Integrated Cell-Material Sciences (iCeMS), Kyoto University, 606-8501, Japan

* Corresponding author (dpackwood@icems.kyoto-u.ac.jp)

***Supporting note 1.*** *Structure relaxation settings*

All structure relaxations were performed using density functional theory (DFT) as implemented in the FHI-aims code.

***$NH_2$-MIL-125 interface***

- Exchange correlation functional: Local density approximation (Perdew-Wang parameterization)
- van der Waals corrections: Tkatchenho-Scheffler vdW correction
- k-points grid: 1 1 1
- Relativistic corrections: atomic ZORA
- Relaxation algorithm: trust-radius method with force convergence criterion 0.005 eV/angstrom.

The figure below shows the constraints applied during the structure relaxation. Atoms that were fixed are colored as unshaded circles. Atoms that were free to move are colored as shaded spheres. Grey = carbon, white = hydrogen, red = oxygen, blue = nitrogen, gold = titanium. The box indicated periodic boundaries.

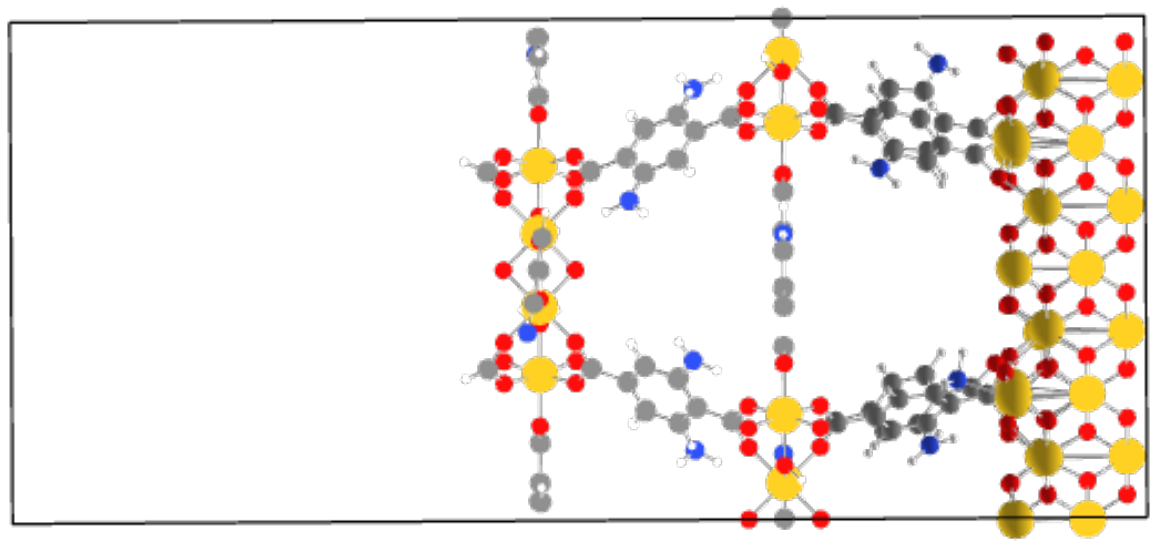

***Cluster and analyte geometry optimization***

- Exchange correlation functional: Local density approximation (Perdew-Wang parameterization)
- van der Waals corrections: Tkatchenho-Scheffler vdW correction
- Relativistic corrections: atomic ZORA
- Relaxation algorithm: trust-radius method with force convergence criterion 0.005 eV/angstrom.

The figure below is showing the constraints applied during the structure relaxation. See the above for details.

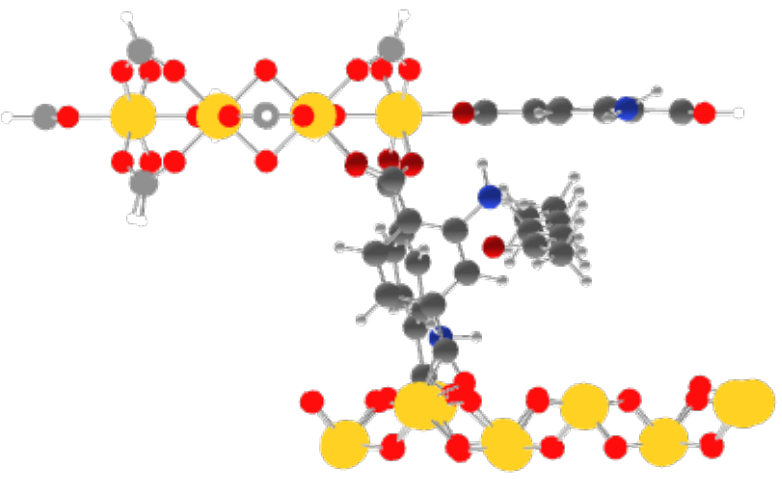

***Supporting note 2. Random docking of molecules inside NH125-$TiO_2$ cluster***

The analyte compounds were placed inside of the cluster as follows:

The pore space is defined by 3 vectors branching out from the metal node of NH2-MIL-125, with each vector aligned with a linker. A random point along each vector is sampled. The random coordinate ($x_r$, $y_r$, $z_r$) is then obtained by projecting these three points into Cartesian coordinates.

The molecule center of mass is translated to the random coordinate ($x_r$, $y_r$, $z_r$). An overlap check between the molecule and the NH125-$TiO_2$ cluster is performed by looping over all atoms in the molecule and all atoms in the NM125-$TiO_2$ cluster. If none of the inter-atomic distances are shorter than the sum of the two covalent radii, the structure is accepted. If not, the process is repeated by sampling another random position.

At the last step, a final overlap check is conducted between the sensor material and the molecule to confirm the integrity of the structure. There are a few cases where the relaxed cluster is not integrable into to the simulation cell of the full interface model, due to the relaxed molecule passing the cluster boundary. In this case, the process is repeated from step ii of Figure 3.

***Supporting Table 1.*** List of 124 compounds found in human breath, their chemical family, binding energy, Wasserstein distance and reference(s).

| CAS | Name | Chemical Family | BE (eV) | WD | Ref |
|---|---|---|---|---|---|
| 106-99-0 | 1,3-butadiene | Alkene | -1.80 | 0.0244 | [1,2] |
| 616-47-7 | 1-methylimidazole | Azoles | -2.99 | 0.0387 | [1] |
| 3391-86-4 | 1-octen-3-ol | Alcohol | -3.78 | 0.0309 | [2,3] |
| 616-25-1 | 1-penten-3-ol | Alcohol | -2.97 | 0.0273 | [4] |
| 431-03-8 | 2,3-butanedione | Ketone | -3.47 | 0.0367 | [2,5] |
| 563-79-1 | 2,3-dimethyl-2-butene | Alkene | -2.34 | 0.0251 | [2,6] |
| 600-14-6 | 2,3-pentanedione | Ketone | -3.12 | 0.0320 | [7] |
| 108-08-7 | 2,4-dimethylpentane | Hydrocarbon | -1.91 | 0.0284 | [2,8] |
| 78-93-3 | 2-butanone | Ketone | -2.35 | 0.0319 | [1,2] |
| 4170-30-3 | 2-butenal | Aldehyde | -2.79 | 0.0304 | [1,2] |
| 110-43-0 | 2-heptanone | Ketone | -2.63 | 0.0249 | [2] |
| 591-78-6 | 2-hexanone | Ketone | -2.92 | 0.0320 | [1,2] |
| 6728-26-3 | 2-hexenal | Aldehyde | -2.13 | 0.0160 | [2] |
| 497-03-0 | 2-methyl-2-butenal | Aldehyde | -2.52 | 0.0195 | [2] |
| 14250-96-5 | 2-methyl-2-pentenal | Hydrocarbon | -2.18 | 0.0285 | [2] |
| 821-55-6 | 2-nonanone | Ketone | -3.52 | 0.0237 | [1,2] |
| 2216-38-8 | 2-nonene | Hydrocarbon | -2.68 | 0.0276 | [2] |
| 2548-87-0 | 2-octenal | Aldehyde | -1.79 | 0.0211 | [2,9] |
| 107-87-9 | 2-pentanone | Ketone | -1.72 | 0.0220 | [2 10] |
| 1576-87-0 | 2-pentenal | Aldehyde | -3.06 | 0.0401 | [11] |
| 112-12-9 | 2-undecanone | Ketone | -4.11 | 0.0343 | [12] |
| 928-80-3 | 3-decanone | Ketone | -3.87 | 0.0296 | [13] |
| 107-86-8 | 3-methyl-2-butenal | Aldehyde | -3.36 | 0.0298 | [2] |
| 589-81-1 | 3-methylheptane | Hydrocarbon | -3.22 | 0.0297 | [2] |
| 589-34-4 | 3-methylhexane | Hydrocarbon | -3.03 | 0.0311 | [2] |
| 96-14-0 | 3-methylpentane | Hydrocarbon | -2.71 | 0.0294 | [2,14] |
| 106-68-3 | 3-octanone | Ketone | -3.57 | 0.0338 | [2] |
| 96-22-0 | 3-pentanone | Ketone | -2.72 | 0.0338 | [15] |
| 625-33-2 | 3-penten-2-one | Ketone | -2.39 | 0.0292 | [2] |
| 2216-87-7 | 3-undecanone | Hydrocarbon | -3.12 | 0.0320 | [2,16] |
| 123-19-3 | 4-heptanone | Ketone | -2.52 | 0.0277 | [2] |
| 108-10-1 | 4-methyl-2-pentanone | Ketone | -2.56 | 0.0225 | [2] |
| 589-53-7 | 4-methylheptane | Hydrocarbon | -5.23 | 0.0322 | [2] |
| 75-07-0 | Acetaldehyde | Aldehyde | -1.47 | 0.0199 | [17,18] |
| 71-50-1 | Acetate | Acid | -5.21 | 0.0466 | [1] |
| 67-64-1 | Acetone | Ketone | -1.69 | 0.0309 | [1,2] |
| 98-86-2 | Acetophenone | Ketone | -2.65 | 0.0226 | [2] |
| 107-02-8 | Acrolein | Aldehyde | -2.67 | 0.0290 | [1,2] |
| 7664-41-7 | Ammonia | Gas | -0.80 | 0.0211 | [1,2] |
| 100-52-7 | Benzaldehyde | Aldehyde | -1.96 | 0.0216 | [19] |
| 71-43-2 | Benzene | Aromatic | -1.91 | 0.0264 | [2] |
| 123-72-8 | Butanal | Aldehyde | -2.40 | 0.0232 | [1,2] |
| 106-97-8 | Butane | Hydrocarbon | -1.36 | 0.0275 | [1,2] |
| 64-19-7 | CH3COOH | Acid | -2.24 | 0.0290 | [2] |
| 124-38-9 | CarbonDioxide | Gas | -0.62 | 0.0252 | [1,2] |
| 630-08-0 | CarbonMonoxide | Gas | -1.36 | 0.0205 | [1,2] |
| 2280-44-6 | D-Glucose | Lipid | -3.43 | 0.0297 | [1,2] |
| 75-18-3 | DimethylSulfide | Sulphide | -1.33 | 0.0215 | [1,2] |
| 124-40-3 | Dimethylamine | Amine | -1.94 | 0.0194 | [20,21] |
| 143-07-7 | DodecanoicAcid | Acid | -4.00 | 0.0364 | [22,23] |
| 64-17-5 | Ethanol | Alcohol | -1.28 | 0.0248 | [1,2] |
| 141-78-6 | EthylAcetate | Acetate | -2.20 | 0.0321 | [2] |
| 100-41-4 | Ethylbenzene | Aromatic | -2.37 | 0.0274 | [1,2] |
| 50-00-0 | Formaldehyde | Aldehyde | -3.64 | 0.0324 | [2] |
| 71-47-6 | Formate | Acid | -5.10 | 0.0492 | [1] |
| 111-71-7 | Heptanal | Aldehyde | -2.57 | 0.0257 | [1,2] |
| 111-14-8 | HeptanoicAcid | Lipid | -3.10 | 0.0289 | [2] |
| 66-25-1 | Hexanal | Aldehyde | -2.95 | 0.0231 | [1,2] |
| 142-62-1 | HexanoicAcid | Lipid | -3.16 | 0.0387 | [1,2] |
| 111-27-3 | Hexanol | Alcohol | -2.24 | 0.0222 | [1] |
| 7722-84-1 | HydrogenPeroxide | Radical | -1.60 | 0.0288 | [1,2] |
| 7786-06-4 | HydrogenSulfide | Sulphide | -1.04 | 0.0271 | [2,24,25] |
| 7803-49-8 | Hydroxylamine | Ammonium | -3.17 | 0.0247 | [26] |
| 120-72-9 | Indole | Aromatic | -2.96 | 0.0359 | [27] |
| 79-31-2 | IsobutyricAcid | Acid | -2.51 | 0.0250 | [2] |
| 78-78-4 | Isopentane | Hydrocarbon | -2.03 | 0.0273 | [2] |
| 78-79-5 | Isoprene | Hydrocarbon | -2.17 | 0.0244 | [1,2,28] |
| 67-63-0 | Isopropanol | Alcohol | -2.15 | 0.0310 | [1,2] |
| 542-78-9 | Malonaldehyde | Aldehyde | -1.92 | 0.0317 | [1,2] |
| 14493-06-2 | Methane | Hydrocarbon | -0.62 | 0.0232 | [2] |
| 74-93-1 | Methanethiol | Sulphide | -2.01 | 0.0273 | [1,2] |
| 67-56-1 | Methanol | Alcohol | -1.07 | 0.0233 | [1,2] |
| 1634-04-4 | MethylTert-butylEther | Alcohol | -2.12 | 0.0276 | [2] |
| 108-88-3 | Methylbenzene | Aromatic | -2.73 | 0.0325 | [2] |
| 10102-43-9 | NitricOxide | Radical | -0.65 | 0.0356 | [2] |
| 124-19-6 | Nonanal | Aldehyde | -0.61 | 0.0335 | [2,14,29] |
| 124-13-0 | Octanal | Aldehyde | -2.49 | 0.0261 | [1,2] |
| 124-07-2 | OctanoicAcid | Acid | -3.55 | 0.0332 | [2] |
| 110-62-3 | Pentanal | Aldehyde | -2.99 | 0.0317 | [1,2] |
| 108-95-2 | Phenol | Aromatic | -2.96 | 0.0274 | [1,2] |
| 74-98-6 | Propane | Hydrocarbon | -1.45 | 0.0232 | [1,2] |
| 71-23-8 | Propanol | Alcohol | -1.48 | 0.0201 | [1,2,30] |
| 110-86-1 | Pyridine | Aromatic | -2.76 | 0.0318 | [2,28] |
| 100-42-5 | Styrene | Aromatic | -2.54 | 0.0271 | [1,2] |
| 7446-09-5 | SulfurDioxide | Sulphide | -2.23 | 0.0284 | [1,2] |
| 108-88-3 | Toluene | Aromatic | -2.30 | 0.0279 | [2,31,32] |
| 75-50-3 | Trimethylamine | Amine | -1.93 | 0.0229 | [2,33] |
| 109-52-4 | ValericAcid | Acid | -2.96 | 0.0243 | [2,34] |
| 106-42-3 | Xylene | Aromatic | -2.69 | 0.0301 | [2] |
| 75-65-0 | tert-Butanol/ 2-methyl-2-propanol | Alcohol | -2.59 | 0.0233 | [2] |
| 526-73-8 | 1,2,3-trimethylbenzene | Aromatic | -2.77 | 0.0277 | [2] |
| 95-63-6 | 1,2,4-trimethylbenzene | Aromatic | -2.45 | 0.0251 | [2] |
| 592-76-7 | 1-heptene | Hydrocarbon | -2.78 | 0.0292 | [2] |
| 62199-20-6 | 2,2,6,6-Tetramethyloctane | Hydrocarbon | -1.97 | 0.0344 | [35] |
| 2213-23-2 | 2,4-Dimethylheptane | Hydrocarbon | -2.98 | 0.0310 | [2] |
| 1072-05-5 | 2,6-Dimethylheptane | Hydrocarbon | -3.52 | 0.0333 | [2] |
| 111-76-2 | 2-Butoxyethanol | Alcohol | -3.65 | 0.0274 | [2] |
| 3221-61-2 | 2-Methyloctane | Hydrocarbon | -2.80 | 0.0277 | [2] |
| 98-83-9 | 2-Phenylpropene | Aromatic | -3.20 | 0.0294 | [1] |
| 2216-33-3 | 3-methyloctane | Hydrocarbon | -2.92 | 0.0292 | [2] |
| 17301-32-5 | 4,7-Dimethylundecane | Hydrocarbon | -0.07 | 0.0355 | [2] |
| 2216-34-4 | 4-methyloctane | Hydrocarbon | -3.35 | 0.0288 | [2] |
| 1632-70-8 | 5-Methylundecane | Hydrocarbon | -3.12 | 0.0280 | [2] |
| 598-58-3 | Methylnitrate | Acid | -1.91 | 0.0219 | [36,37] |
| 626-38-0 | 1-Methylbutylacetate | Acid | -3.83 | 0.0372 | [2] |
| 513-86-0 | Acetoin | Ketone | -1.97 | 0.0222 | [2] |
| 98-86-2 | Acetophenone | Ketone | -2.22 | 0.0225 | [2] |
| 107-13-1 | acrylonitrile | Nitrile | -2.18 | 0.0296 | [2] |
| 100-47-0 | Benzonitrile | Nitrile | -2.02 | 0.0258 | [2] |
| 463-58-1 | Carbonylsulfide | Sulphide | -0.71 | 0.0259 | [2] |
| 110-82-7 | Cyclohexane | Hydrocarbon | -1.98 | 0.0270 | [2] |
| 124-18-5 | Decane | Hydrocarbon | -4.15 | 0.0300 | [2] |
| 75-09-2 | Dichloromethane | Hydrocarbon | -1.60 | 0.0257 | [2] |
| 74-84-0 | Ethane | Hydrocarbon | -1.00 | 0.0247 | [2] |
| 23676-09-7 | Ethyl-4-ethoxybenzoate | Acid | -4.52 | 0.0377 | [38,39] |
| 98-01-1 | Furfural | Aldehyde | -2.55 | 0.0203 | [2] |
| 142-82-5 | Heptane | Hydrocarbon | -1.60 | 0.0278 | [2] |
| 96-37-7 | Methylcyclopentane | Hydrocarbon | -2.10 | 0.0262 | [2] |
| 95-53-4 | O-toluidine | Aromatic | -2.74 | 0.0342 | [40,41] |
| 111-65-9 | Octane | Hydrocarbon | -3.24 | 0.0272 | [2] |
| 109-66-0 | Pentane | Hydrocarbon | -2.08 | 0.0255 | [2] |
| 123-38-6 | Propanal | Aldehyde | -1.50 | 0.0297 | [2] |
| 103-65-1 | Propyl benzene | Aromatic | -2.79 | 0.0286 | [42] |
| 1120-21-4 | Undecane | Hydrocarbon | -4.10 | 0.030<br>1 | [2] |

***Supporting Table 2.*** List of disease biomarkers collected from selected studies.[1,43,44]

| | |
|---|---|
| COPD | Acetate<br>Heptanal<br>Hexanal<br>Malondialdehyde<br>Nonanal<br>Isoprene<br>NitricOxide<br>2,4,6-Trimethyldecane<br>2,6-Dimethylheptane<br>4,7-Dimethylundecane<br>4-methyloctane<br>Ethane<br>Hexadecane<br>Octadecane<br>Undecane<br>Benzonitrile |
| Diabetes | D-Glucose<br>Acetone<br>Methylnitrate |
| Hepatic Disease | 2-pentanone<br>2-butanone<br>1-octen-3-ol<br>DimethylSulfide<br>Acetone<br>Limonene<br>Methanol |
| Liver cancer | Propanoic acid<br>1-Hexadecanol<br>Isopropanol<br>Acetaldehyde<br>Octane<br>2-pentanone<br>Acetone<br>Carbonylsulfide<br>Dimethyl sulphide<br>Methanethiol |
| Pancreas Disease | Trimethylamine<br>Ammonia |
| Gastrointestinal Cancer | Furfural<br>4,5-Dimethylnonane<br>4-methyloctane<br>Hexadecane<br>Isoprene<br>1,2,3-trimethylbenzene<br>2-Phenylpropene<br>1-Methylbutylacetate<br>2-Butoxyethanol<br>2-butanone<br>6-methyl-5-hepten-2-one<br>acrylonitrile |
| Gastrointestinal disease | HexanoicAcid<br>Hexanal<br>Heptanal<br>Nonanal<br>Pentanal<br>Phenol<br>Octanal<br>Butanal |
| Renal disease | Ammonia<br>Isopropanol<br>2,4-Dimethylheptane<br>2-Methyloctane<br>Nonane<br>Ethylbenzene<br>Styrene<br>Dichloromethane |
| Kidney Injury | Ethanol<br>2-pentanone<br>Acetone |
| Chronic Renal Failure (Uremia) | Ammonia<br>2-butanone<br>Dimethylamine<br>Trimethylamine<br>2,2,6,6-Tetramethyloctane |
| | 2,4-Dimethylheptane |
| Lung Cancer | Ethanol<br>Methanol<br>Butanal<br>Formaldehyde<br>Heptanal<br>Hexanal<br>Nonanal<br>Octanal<br>Pentanal<br>Propanal<br>3-methylhexane<br>3-methyloctane<br>Butane<br>Cyclohexane<br>Decane<br>Heptane<br>Methylcyclopentane<br>Octane<br>Pentane<br>Undecane<br>1-heptene<br>Isoprene<br>1,2,4-trimethylbenzene<br>2,3-dihydro-1,1,3-trimethyl-3- phenyl-1-H-indene<br>2,5-dimethyl furan<br>Benzene<br>Ethyl-4-ethoxybenzoate<br>O-toluidine<br>Propylbenzene<br>Styrene<br>Toluene<br>Acetoin<br>Acetone |
| Multiple Sclerosis | Decanal<br>Hexanal<br>Nonanal<br>5-Methylundecane |
| | Heptadecane<br>Acetophenone<br>Sulfur dioxide |

***Supporting Table 3.*** List of nitroexplosive and interfering compounds

| CAS | Name | Group | BE (eV) | Wd |
|---|---|---|---|---|
| 528-29-0 | O-DNB / O-Dinitrobenzene | Nitroexplosive | -5.37 | 0.0532 |
| 121-82-4 | RDX / Cyclonite | Nitroexplosive | -3.76 | 0.0314 |
| 72454-49-0 | TNP / 2,4,6-Trinitrophenol | Nitroexplosive | -5.33 | 0.0692 |
| 118-96-7 | TNT / 2,4,6-Trinitrotoluene | Nitroexplosive | -5.64 | 0.0614 |
| 67-64-1 | Acetone | Interferents | -1.69 | 0.0309 |
| 7664-41-7 | Ammonia | Interferents | -0.80 | 0.0211 |
| 71-43-2 | Benzene | Interferents | -1.91 | 0.0264 |
| 124-38-9 | Carbon Dioxide | Interferents | -0.62 | 0.0252 |
| 630-08-0 | Carbon Monoxide | Interferents | -1.36 | 0.0205 |
| 64-17-5 | Ethanol | Interferents | -1.28 | 0.0248 |
| 7783-06-4 | Hydrogen Sulfide | Interferents | -1.04 | 0.0271 |
| 14493-06-2 | Methane | Interferents | -0.62 | 0.0232 |
| 7446-09-5 | Sulfur Dioxide | Interferents | -2.23 | 0.0284 |
| 108-88-3 | Toluene | Interferents | -2.30 | 0.0279 |

***Supporting Note 3.* Calculation of autocorrelation functions**

Autocorrelation functions (ACFs) were computed as

$$\rho(k) = \frac{\langle (g(X_0) - \mu_0)(g(X_k) - \mu_k) \rangle}{\sigma_0 \sigma_k},$$

where

$$\mu_k = \langle g(X_k) \rangle$$

and

$$\sigma_k = \left\langle (g(X_k) - \mu_k)^2 \right\rangle^{1/2}.$$

In the above, $X$ represents a random walk (Markov chain) which hops between points in the principal component plot. $X_k$ is the point at which the random walk resides after $k$ hops. $g(X_k)$ is the value of the property of interest (binding energy, Wasserstein distance, or random noise) for this point. The angular brackets represent the expected value. The random walk was simulated such that, for a given point in the principal component plot, a hop was permitted to any neighboring point within a cut-off radius of 25.0 with equal probability. The ACFs were computed by simulating 10,000 random walks independently of each other, each initiated from a random initial point.

***Supporting Figure 1.*** Wasserstein distance calculation

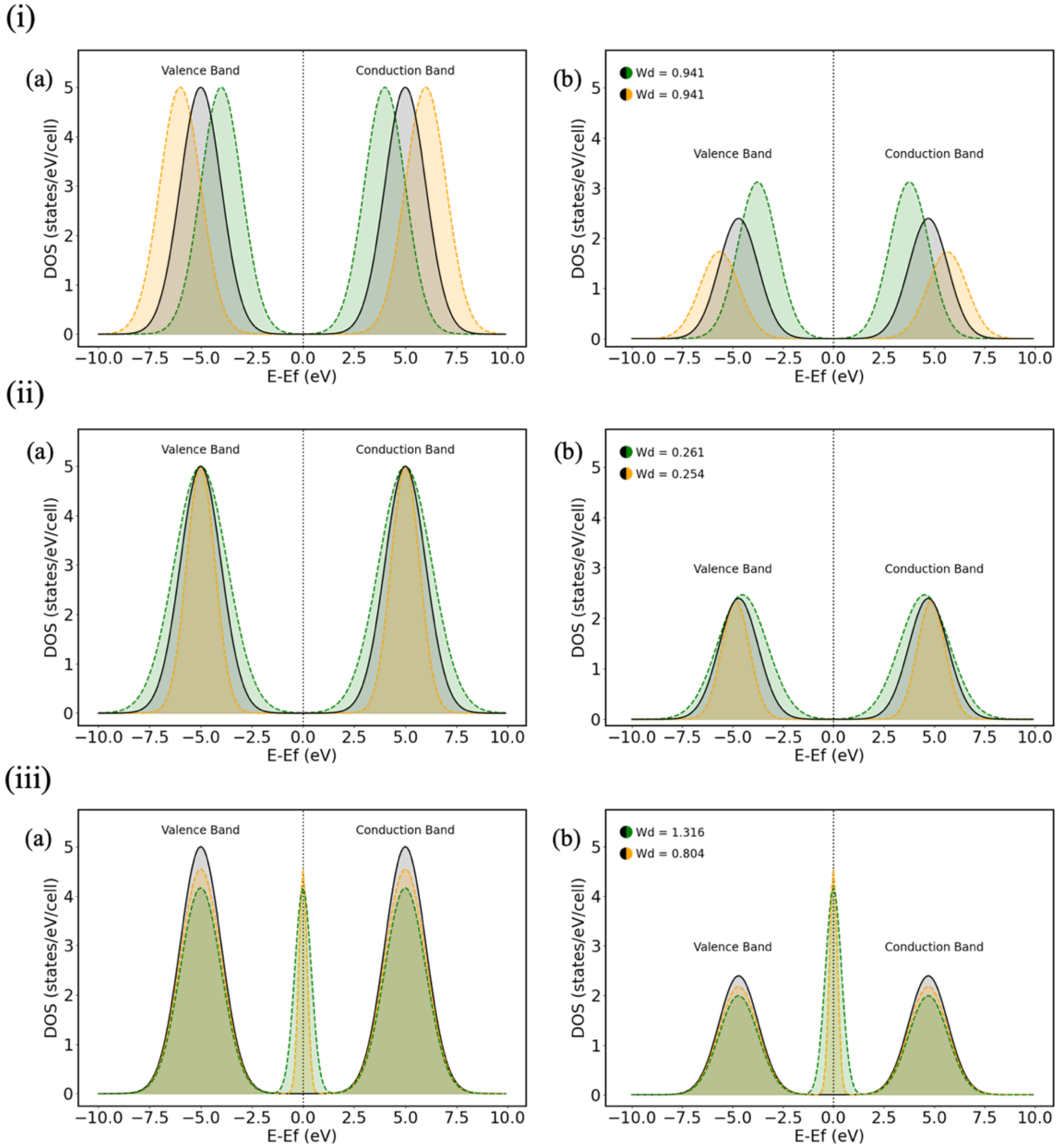

Three hypothetic density of states (DOS) curves. The black dotted line is the fermi level, while the black solid line forming two gaussian represents a conduction and a valence band of a DOS profile. (a) The conduction and valence band move closer (green dotted line) and further away (orange dotted line) from the fermi level. (b) Wider (green dotted line) and narrower (orange dotted line) conduction band and valence band. (f) A third gaussian representing a new band is formed on the fermi level. Gaussian weighting is applied on (i.a, ii.a, iii.a) DOS profiles resulting to (i.b, ii.b, iii.b) DOS profiles, respectively. The Wasserstein distance (Wd) between the green DOS and the black DOS, and the orange DOS and the black DOS are calculated.

***Supporting references***